\documentclass[12pt]{article}
\usepackage{amsfonts}
\usepackage{amssymb}
\usepackage{graphics,psboxit,amsmath}
\usepackage{subfigure}
\usepackage{graphicx}
\usepackage{verbatim}
\usepackage{epic}

\def\hybrid{\topmargin 0pt \oddsidemargin 0pt 
        \headheight 0pt \headsep 0pt
        \textwidth 16,5cm 
        \textheight 23cm 
        \marginparwidth .875in
        \parskip 5pt plus 1pt \jot = 1.5ex}

\hybrid

\catcode`\@=11

\def\marginnote#1{}
\newcount\hour
\newcount\minute
\newtoks\amorpm
\hour=\time\divide\hour by60
\minute=\time{\multiply\hour by60 \global\advance\minute by-\hour}
\edef\standardtime{{\ifnum\hour<12 \global\amorpm={am}%
        \else\global\amorpm={pm}\advance\hour by-12 \fi
        \ifnum\hour=0 \hour=12 \fi
        \number\hour:\ifnum\minute<10 0\fi\number\minute\the\amorpm}}
\edef\militarytime{\number\hour:\ifnum\minute<10 0\fi\number\minute}

\def\draftlabel#1{{\@bsphack\if@filesw {\let\thepage\relax
   \xdef\@gtempa{\write\@auxout{\string
      \newlabel{#1}{{\@currentlabel}{\thepage}}}}}\@gtempa
   \if@nobreak \ifvmode\nobreak\fi\fi\fi\@esphack}
        \gdef\@eqnlabel{#1}}
\def\@eqnlabel{}
\def\@vacuum{}
\def\draftmarginnote#1{\marginpar{\raggedright\scriptsize\tt#1}}

\def\draft{\oddsidemargin -.5truein
        \def\@oddfoot{\sl preliminary draft \hfil
        \rm\thepage\hfil\sl\today\quad\militarytime}
        \let\@evenfoot\@oddfoot \overfullrule 3pt
        \let\label=\draftlabel
        \let\marginnote=\draftmarginnote
   \def\@eqnnum{(\theequation)\rlap{\kern\marginparsep\tt\@eqnlabel}%
\global\let\@eqnlabel\@vacuum} }

\def\draft2{
        \def\@oddfoot{\sl preliminary draft \hfil
        \rm\thepage\hfil\sl\today\quad\militarytime}
        \let\@evenfoot\@oddfoot \overfullrule 3pt
        \let\label=\draftlabel
        \let\marginnote=\draftmarginnote
   \def\@eqnnum{(\theequation)\rlap{\kern\marginparsep\tt\@eqnlabel}%
\global\let\@eqnlabel\@vacuum} }

\def\preprint{\twocolumn\sloppy\flushbottom\parindent 2em
        \leftmargini 2em\leftmarginv .5em\leftmarginvi .5em
        \oddsidemargin -.5in \evensidemargin -.5in
        \columnsep .4in \footheight 0pt
        \textwidth 10.in \topmargin -.4in
        \headheight 12pt \topskip .4in
        \textheight 6.9in \footskip 0pt
        \def\@oddhead{\thepage\hfil\addtocounter{page}{1}\thepage}
        \let\@evenhead\@oddhead \def\@oddfoot{} \def\@evenfoot{} }

\def\numberbysection{\@addtoreset{equation}{section}
        \def\theequation{\thesection.\arabic{equation}}}

\def\underline#1{\relax\ifmmode\@@underline#1\else
        $\@@underline{\hbox{#1}}$\relax\fi}

\def\titlepage{\@restonecolfalse\if@twocolumn\@restonecoltrue\onecolumn
     \else \newpage \fi \thispagestyle{empty}\c@page\z@
        \def\thefootnote{\fnsymbol{footnote}} }

\def\endtitlepage{\if@restonecol\twocolumn \else \newpage \fi
        \def\thefootnote{\arabic{footnote}}
        \setcounter{footnote}{0}} 

\catcode`@=12
\relax

\def\figcap{\section*{Figure Captions\markboth
        {FIGURECAPTIONS}{FIGURECAPTIONS}}\list
        {Figure \arabic{enumi}:\hfill}{\settowidth\labelwidth{Figure
999:}
        \leftmargin\labelwidth
        \advance\leftmargin\labelsep\usecounter{enumi}}}
 \relax
\def\tablecap{\section*{Table Captions\markboth
        {TABLECAPTIONS}{TABLECAPTIONS}}\list
        {Table \arabic{enumi}:\hfill}{\settowidth\labelwidth{Table
999:}
        \leftmargin\labelwidth
        \advance\leftmargin\labelsep\usecounter{enumi}}}
 \relax
\def\reflist{\section*{References\markboth
        {REFLIST}{REFLIST}}\list
        {[\arabic{enumi}]\hfill}{\settowidth\labelwidth{[999]}
        \leftmargin\labelwidth
        \advance\leftmargin\labelsep\usecounter{enumi}}}
 \relax
\makeatletter
\newcounter{pubctr}
\def\publist{\@ifnextchar[{\@publist}{\@@publist}}
\def\@publist[#1]{\list
        {[\arabic{pubctr}]\hfill}{\settowidth\labelwidth{[999]}
        \leftmargin\labelwidth
        \advance\leftmargin\labelsep
        \@nmbrlisttrue\def\@listctr{pubctr}
        \setcounter{pubctr}{#1}\addtocounter{pubctr}{-1}}}
\def\@@publist{\list
        {[\arabic{pubctr}]\hfill}{\settowidth\labelwidth{[999]}
        \leftmargin\labelwidth
        \advance\leftmargin\labelsep
        \@nmbrlisttrue\def\@listctr{pubctr}}}
 \relax
\makeatother

\def\ba{\begin{equation}}
\def\ea{\end{equation}}

\def\a{\alpha}

\def\b{\beta}

\def\d{\delta}

\def\e{\epsilon}

\def\no{\noindent}

\def\IR{\relax{\rm I\kern-.18em R}}

\begin{document}


\renewcommand{\theequation}{\thesection.\arabic{equation}}
\csname @addtoreset\endcsname{equation}{section}

\newcommand{\eqn}[1]{(\ref{#1})}
\newcommand{\be}{\begin{eqnarray}}
\newcommand{\ee}{\end{eqnarray}}
\newcommand{\non}{\nonumber}
\begin{titlepage}
\strut\hfill
\begin{center}

\vskip -1 cm


\vskip 2 cm

{\Large \bf  Toward a  standard model 2,  via Kaluza ansatz 2}

\vskip .9 in

{\bf Nikolaos A. Batakis}

\vskip 0.2in

Department of Physics, University of Ioannina, \\
45110 Ioannina,  Greece\\
{\footnotesize{\tt nbatakis@uoi.gr}}\\

\end{center}

\vskip .4in

\centerline{\bf Abstract}

\no
New results and  perspectives  precipitate
from the (modified as) Kaluza ansatz 2 (KA2), whereby,
instead of appending $n$ Planck-scale (${\rm L_o}$) compact SL
dimensions  to  ordinary 4D spacetime, one {\em augments} $n$
such dimensions by 3 large ones. By KA2,  the  fundamental  
r\^ole of gravity in the dynamics of vacuum geometry is being 
conceded to the remaining fundamental interactions.
The ground state in KA2 is of the form 
$\bar{\cal M}^{n+4}\!=\!\bar{\cal C}^{n+1}\!\times\!\IR^3$,
where the static (averaged-out over scales  
${\rm L}\!>\!>\!{\rm L_o}$) $\bar{\cal C}^{n+1}$
carries {\em effective torsion} 
as  relic of  the deeper vacuum dynamics at Planck scale. 
For the simplest non-trivial implementation of  KA2, 
the Bianchi IX subclass of $SU(2)$-invariant ${\cal B}^4_{\rm IX}$
provides the $\bar{\cal C}^5\!=\!\bar{\cal B}^4_{\rm M}\times S^1$,
with the $S^1$ coming from `augmentability',
a complement to compactification.
The classical action involves   
(i) the gravitational and EW  sectors in elegant {\em hierarchy},
(ii) the {\em higgsless}  emergence and full calculability of the EW gauge bosons masses
and (iii) gravity as a necessarily 
effective field, hence  non-quantizable. 
A conjectured ${\cal C}^{n+1}$  with $n\leq 7$ (to adjoin the 
strong interaction) toward a standard model 2,  might also
offer novel perspectives for supergravity.

\vskip .3 in
\no
{\em Keywords:\/}  Kaluza-Klein theories,  Taub string, hierarchy, 
torsion, higgsless EW masses, compactification, augmentability,
standard model, supergravity, quantization of gravity.

\vfill
\no


\end{titlepage}
\vfill
\eject


\section{Introduction}

The main task at LHC may be impeded by the 
Higgs sector of the standard model,
but the latter  will
require  even deeper reform, 
if the former folds at LHC, because of 
its other fundamental problems, notably 
on hierarchy and the quantization of gravity.
Curiously related, forty years before the collective formulation
of the Higgs mechanism, the Kaluza ansatz was likewise received 
as a `clever artifact' (for the enlargement of the then young theory
of general relativity), to be  likewise elevated subsequently 
to a fundamental notion,  but its geometric elegance 
has remained unquestionably unique all along. The
fundamental interactions can be segregated by physical aspects
(dimension-less vs -full couplings and quantization) but not by
{\em a priori} geometrical ones in a unified
higher-dimensional  context. Nevertheless,
one may resort to the {\em approach} towards full  geometrization via the
standard Kaluza ansatz  \cite{duff}, for a complementary
KA2 approach, in the sense that, 
instead of appending $n$ Planck-scale (${\rm L_o}$) compact SL
dimensions  to  ordinary general-relativistic 4D spacetime, 
one can {\em augment} $n$
such dimensions of a ${\cal C}^{n+1}$ 
proper vacuum \cite{bat1} by 3 large ones.

By KA2,  the central  
r\^ole of gravity in the dynamics of vacuum geometry is being 
conceded to the remaining fundamental interactions.
The ground state in KA2  is of the form 
$\bar{\cal M}^{n+4}\!=\!\bar{\cal C}^{n+1}\!\times\!\IR^3$,
where the static (averaged-out over scales  
${\rm L}\!>\!>\!{\rm L_o}$) $\bar{\cal C}^{n+1}$ carries
{\em effective torsion} \cite{bat1}, \cite{traut} 
as  relic of  the deeper vacuum dynamics at Planck scale. 
The requirement of {\em augmentability}
(a complement to that of  spontaneous compactification)  will
curtail the already-small class of ${\cal C}^{n+1}$.
The latter can be a homogeneous space  \cite{helga}, \cite{rs}   
with vigorous or even chaotic dynamics, 
like  Misner's elusive mixmaster  ${\cal B}^4_{\rm M}$ \cite{misner}
in the  Bianchi IX subclass of  left-$SU(2)$ invariant 
${\cal B}^4_{\rm IX}$, which also includes
the Taub string \cite{bat1}, a pp wave 
${\cal B}^4_{\rm T}$. In the static $\bar{\cal B}^4_{\rm M}$, 
$\bar{\cal B}^4_{\rm T}$, the effective torsion $\bar T^A$
(parallelizing in the second case)  is explicitly
calculable via the  effective  loss of  the  Ricci flatness in
${\cal B}^4_{\rm M}$, 
${\cal B}^4_{\rm T}$ \cite{bat1}. 
The simplest possible non-trivial ground state for K-A2
involves a $\bar{\cal M}^8$ with a $\bar{\cal C}^5\!=
\!\bar{\cal B}^4_{\rm M}\times S^1$, where
$\bar{\cal B}^4_{\rm M}$ is chosen for its 
round rather than squashed $S^3$; the $S^1$
factor is actually {\em imposed} 
by  augmentability, as we will see, whereby the topology of the  SL sections and
the transitive $SU(2)$ invariance on them
must  be enlarged to at-least $S^3\!\times\!S^1$ and 
$SU(2)\!\times\!U(1)$, respectively. 
For a comparative view, we will also cite the standard
$M^8_{\rm o}$ ground state (with Minkowski's $M^4_{\rm o}$)
as depicted in the schemes 
\be
M^8_{\rm o}(\bar e^A, \bar\Gamma^A_{\;\;B}):\;=
M^4_{\rm o}\times \left(S^3\times S^1\right)\;\;\;\;\;\;\;
\stackrel{\delta\bar e}{\longrightarrow}
\;\;\;\;\;\;\;\; M^8(e^A,\Gamma^A_{\;\;B})
\label{m8s}
\ ,
\ee
\be
\bar{\cal M}^8(\bar e^A, \bar\gamma^A_{\;\;B}):\,=
\left(\bar{\cal B}^4_{\rm M}\times S^1\right)\times \IR^3\;\;\;\;\;\;
\stackrel{\delta\bar e,\,\delta\bar\gamma}{\longrightarrow}
\;\;\;\;\;\;\;{\cal M}^8(e^A, \gamma^A_{\;\;B})
\ ,
\label{m8}
\ee
for the standard vs the  KA2 approach.
The difference between $M^8_{\rm o}$ and 
$\bar {\cal M}^8$ may at first appear to be rather trivial,
because they both have  the same topology and 
metric (or $\bar e^A$ Cartan frames), hence also identical 
$\bar\Gamma^A_{\;\;B}$ Christoffel connections, so their
only difference is the presence of the effective torsion
$\bar T^A$ in the $\bar\gamma^A_{\;\;B}$ 
connection of $\bar {\cal M}^8$. 
Nevertheless, their difference  in perspective
and results will turn out to be fundamental.
In either case  we arrive
at a `low-energy'
configuration, the $ M^8(e^A,\Gamma^A_{\;\;B})$,
${\cal M}^8(e^A, \gamma^A_{\;\;B})$, respectively.

In the standard case, the process of
starting with the $M^8_{\rm o}(\bar e^A, \bar\Gamma^A_{\;\;B})$ 
ground state to arrive  by (\ref{m8s}) at  $ M^8(e^A,\Gamma^A_{\;\;B})$ is
geometrically viewable as a  tilt from excitation of the frames 
from their value $\bar e^A$ in $M^8_{\rm o}$
to  $ e^A=\bar e^A+\delta\bar e^A$ in $ M^8(e^A,\Gamma^A_{\;\;B})$.
The physical content of this excitation is, of course, the $SU(2)\times U(1)$
gauge-field potentials  ${\cal A}^I$, which,
in the case of (\ref{m8}) for the  KA2 approach, must {\em also} excite the
torsion $\bar T^A$ in $\bar{\cal M}^8$. 
By the holonomy theorems and the Cartan structure equations for any 
(${\cal R}^A_{\;\;B}$, $T^A$) set \cite{traut},
the torsion $T^A$ (field-content {\em and} scale) is completely independent 
from the Riemannian part $ R^A_{\;\;B}$ of the curvature ${\cal R}^A_{\;\;B}$. 
Thus,  excitations under the  KA2 approach to reach
${\cal M}^8(e^A, \gamma^A_{\;\;B})$  in (\ref{m8})
must be of the `metric {\em and}
connection'  Palatini type, namely independent
excitations of  frames  and of torsion, so they will necessarily involve 
(beyond $\kappa_{\rm o}$, ${\rm L_o}$)
two new independent scales, 
the $\kappa$ and ${\rm L_1}$, 
respectively. Classically they can be of virtually any amplitude,
limited only by the strength $\kappa_{\rm o}^{-2}$ 
of the Taub-string, which is of  Planck scale, 
and likewise for ${\cal B}^4_{\rm M}$.
However, as with the otherwise stable Minkowski 
vacuum in the standard approach, the  addition `by-hand'
of  {\em any}  mass 
in ${\cal B}^4_{\rm T}$ or ${\cal B}^4_{\rm M}$  
would cause a mathematical singularity \cite{rs}.
As long as this cannot be averted by the overlying torsion, 
mass terms in the respective actions  can
be  generated only {\em effectively} by the  geometry,
or the  vacuum stability will be lost.

\no
\underline {Notes on notation:}
The indices $A,B,M\dots$ run as
$M\!=\!(\mu;m)\!=\!(0,5,6,7;1,2,3,4)$  with (1,2,3,4) in the compact dimensions.
 In all our Cartan frames and duals 
($e^M\!,\, E_N$)  we employ orthonormal $\eta_{AB}\!=\!{\rm diag} (-1,+1,\dots,+1)$
and $I\!=\!(i,4)\!=\!(1,2,3,4)$ indices for the $SU(2)\!\times\!U(1)$  left-invariant  1-forms with
$d\ell^i\!=\!-\frac{1}{2}\epsilon^i_{jk}\ell^j\ell^k$, $d\ell^4\!=\!0$.
Due to isometries on $S^3\times S^1$
there exist four transitive Killing vectors $\Xi_I$  
and their components $\Xi_I^m$ remain invariant
under both types of the Kaluza ansatz. Commutation relations between the 
$L_\mu$, $\Xi_\nu$ and  Lie derivatives ${\cal{L}}_{\Xi_I}$ 
(by use of the duality relation $\ell^m(L_n)\!=\!\delta^m_n$, etc.) 
can be summarized as \cite{duff}
\be
[L_j,L_m]=\delta^k_m \epsilon^i{\!}_{jk}{\,}L_i,\;\;\;\;
[\Xi_j,\Xi_m]= \delta^k_m\epsilon^i{\!}_{jk}{ }\Xi_i ,\;\;\;\;
{\cal{L}}_{\Xi_I}L_m:=[\Xi_I,L_m]=0,\;\;\;\;\;{\cal{L}}_{\Xi_I}\ell^m=0.
\ 
\label{com}
\ee
The general connection $\gamma^M_{\;\;N}$ and the Christoffel 
$\Gamma^M_{\;\;N}\!=\Gamma^M_{\;\;NP}e^P$
in the covariant derivatives ${\cal D}$, $D$,  respectively,  
are antisymmetric in $M,N$ just like the
contorsion tensor-valued 1-form $K^M_{\;\;N}$ in
$\gamma^M_{\;\;N}=\Gamma^M_{\;\;N} +K^M_{\;\;N}$ with
$De^M:=de^M+\Gamma^M_{\;\;N}\wedge e^N\equiv 0$,
$DE_M=dE_M-\Gamma^N_{\;\;M}E_N\equiv 0$.
The general curvature ${\cal R}^A_{\;\;B}$
includes its Riemannian  part $R^A_{\;\;B}:=
d\Gamma^A_{\;\;B}+\Gamma^A_{\;\;P}\wedge
\Gamma^P_{\;\;B}$, with the Weyl and Ricci tensors
${W}^A_{\;\;BMN}$  and
$R_{MN}\!=\!R^P_{\;MPN}$ .
Cartan's first and second structure equations 
involve the general curvature ${\cal R}^M_{\;\;N}$
and the torsion $T^M$ 2-forms  as  \cite{traut}
\be
{\cal R}^A_{\;\;B}\,:\!&=&\!d\gamma^A_{\;\;B}+\gamma^A_{\;\;P}
\wedge\gamma^P_{\;\;B}=R^A_{\;\;B}+
DK^A_{\;\;B}+K^A_{\;\;P}\wedge K^P_{\;\;B}=
\frac{1}{2}{\cal R}^A_{\;\;BNP}\,e^N\wedge e^P ,
\label{riemann}\\
T^M\,:\!&=&\!{\cal D}e^M=de^M+\gamma^M{\!}_N\wedge e^N =
K^M{\!}_N\wedge e^N=
\frac{1}{2}\,T^M_{\;\;NP}\, e^N\wedge e^P 
\label{torsion}
\ . 
\ee


\no
\section{Tilting the frames in
$\bar{\cal M}^8(\bar e^A, \bar\gamma^A_{\;\;B})$
towards ${\cal M}^8(e^A, \gamma^A_{\;\;B})$}

To implement  K-A2, we must fix the frames etc for 
$\bar{\cal M}^8(\bar e^A, \bar\gamma^A_{\;\;B})$
 in (\ref{m8}), then proceed with the tilt $e^A\!=\!\bar e^A+\delta\bar e^A$
in terms of ${\cal A}^I$ and (in the next section) with the excitation $\delta\bar T^A$. 
From $\left(\bar e^M;\,\bar E_N\right)\!=\!
\left(\bar e^\mu\!=\!\delta^\mu_{\bar\mu}\, dx^{\bar\mu},
\,\bar e^m\!=\!{\rm L_o}\ell^m;\; 
\bar E_\nu\!=\!\delta_\nu^{\bar\nu}\,\partial_{\bar\nu},\,\bar E_n\!=
\!{\rm L_o^{-1}}L_n\right)$, with trivial vierbeins $\delta^\mu_{\bar\mu}$
for holonomic $\bar e^\mu$,
we find  $\bar\gamma^A_{\;\;B}\!=\!\bar\Gamma^A_{\;\;B}+\bar K^A_{\;\;B}$ 
and the non-vanishing $\bar\gamma^i_{\;j}\!=\!
2\bar\Gamma^i_{\;j}\!=2\!\bar K^i_{\;j}\!=\!
\epsilon^i_{\;jk}\bar e^k$; 
the  Ricci  and scalar contractions from the Riemannian part 
($\bar R^i_{\;jkl}$) of the full  curvature ($\bar {\cal R}^i_{\;jkl}$)
are $\bar R_{ij}\!=\!1/2{\rm L_o^{-2}}\eta_{ij}$, 
$\bar R\!=\!3/2{\rm L_o^{-2}}$, identical to those of 
$M^8_{\rm o}(\bar e^A, \bar\Gamma^A_{\;\;B})$  in (\ref{m8s}). 
For the Riemann-Cartan geometry in 
$\bar{\cal M}^8_{\rm o}(\bar e^A, \bar\gamma^A_{\;\;B})$ of (\ref{m8}),
the parallelizing torsion gives a  
$\bar K^i_{\;j}\!=\!1/2\epsilon^i_{\;jk}\bar e^k$ in
$\bar{\cal B}^4_{\rm M}$, hence 
$\bar {\cal R}^i_{\;jkl}\!=\!0$. The vanishing 
of the Hilbert-Einstein-Cartan Lagrangian 
$\bar{\cal L}_{\rm HEC}\!=\!\bar {\cal R}$ 
for $\bar{\cal M}^8$ 
offers harmless simplicity until (\ref{gtilt}). 
The orthonormality relations between the  Killing vectors $\Xi_I$
can be expressed in terms of a continuous angle parameter 
$\theta\!\in\!(0,\pi/2)$,  the {\em slicing angle}  
$\theta$ (to be distinguished from
Weinberg's {\it mixing angle}\footnote{
Under gauge symmetry breaking, $\theta$ could be identified
with whatever particular value the $\theta_W$ has.} $\theta_W$),  as 
\be
\Xi_I^m\Xi_J^n\eta_{mn}\!:\,= 
\Big{(} \frac{{\rm L}_o}{\sin{\theta}} \Big{)}^2\eta_{ij}\delta^i_I\delta^j_J+ 
\Big{(} \frac{{\rm L}_o}{\cos{\theta}} \Big{)}^2\eta_{44}\delta^4_I\delta^4_J 
\ .
\label{killinglengths}
\ee
The scale ${\rm L}_o$ of the components is  imposed by the frames $\bar e^A$
and the lengths  ${\rm L}_o/\sin{\theta}$, ${\rm L}_o/\cos{\theta}$ are 
proportional to the radii of $S^3$ and $S^1$ in any particular slicing 
of the $S^3\times S^1$ torus, as fixed by $\theta$.
These $\Xi_I$ provide a  basis
for tangent vectors on $S^3\times S^1$, just like the $L_m$ do.
However, while the  $L_m$  are {\em ab initio} left-invariant, 
by ${\cal{L}}_{\Xi_I}L_m=0$ etc.,
the $\Xi_I$ cannot possibly form a left-invariant basis,
due to the  ${\cal{L}}_{\Xi_I}\Xi_J\neq 0$
relations from (\ref{com}). 
Therefore, under ordinary circumstances, the $\Xi_I$ would be an 
odd and cumbersome (albeit fully 
legitimate) basis to employ in left-invariant environments, 
such as those involving the round or even the squashed   $S^3{\;}$. 
This observation will be useful to us later on.

The excitation of the frames in
$e^A\!=\!\bar e^A+\delta\bar e^A$ (etc., via 
$\ell^m(L_n)\!=\!\delta^m_n$) in  KA2  is  linear in the gauge potentials
${\cal{A}}^I$  and  identical to that of
the standard case  as
\be
\bar e^A\;\rightarrow\;\; e^A\!:\,=\bar e^A+
g\big[\Xi\cdot{\cal{A}}\big]^m\delta_m^A
 \;\;\; \Longleftrightarrow\;\;\;
\bar E_B\;\rightarrow\;\; {E_B}={ }\bar E_B-
g\big[\Xi\cdot{\cal{A}}\big]_\nu\delta^\nu_B
  \ ,
\label{tilt}
\ee
where $g$ is a scaleless coupling parameter\footnote
{The basic scales, like the ${\rm L}_o$,
are carried by the frames. All other quantities  
must have either derivable or inherently independent scale 
(like the ${\rm L}_1$, to be identified with the EW), or be
 scaleless as with, e.g., all entries in (\ref{com}). 
In the scaleless coupling $g=\sqrt{2}\kappa/{\rm L}_o$,
 the denominator de-scales 
$\Xi_I^\alpha$ (circumstantially scaled by ${\rm L}_o$  in 
(\ref{killinglengths})) and  $\kappa$,
identifiable as the $\sqrt{8\pi G_N}$ gravitational coupling,
provides the missing scale.} 
and  the potentials enter through the components of 
the diagonal tensor $\big[\Xi\cdot{\cal{A}}\big]$ of mixed (1,1) 
rank (to be discussed shortly) defined as
\be
\big[\Xi\cdot{\cal{A}}\big]^m\!:\,=
\Xi_i^m{\cal{A}}^i\sin{\theta}+\Xi_4^m{\cal{A}}^4\cos{\theta} ,\;\;\;\;\;\;
\big[\Xi\cdot{\cal{A}}\big]_\nu=
\Xi_i{\cal{A}}^i_\nu\sin{\theta}+\Xi_4{\cal{A}}^4_\nu\cos{\theta} 
  \ . 
\label{dot}
\ee
The transformations in (\ref{tilt}) can be viewed as trivially reversible
with the simple transfer of the terms involving  
$g\big[\Xi\cdot{\cal{A}}\big]$  
on one or the other side of those relations, 
so as to formally also  define  $\bar e^A$   
in terms of ${e^A}$ etc., with the {\em same} components of 
$g\big[\Xi\cdot{\cal{A}}\big]$. This reveals  an underlying {\em tilt invariance},
whereby tensorial components 
like $\Xi_I^m, \;{\cal{A}}_\nu^I,\;
\big[\Xi\cdot{\cal{A}}\big]_\nu^m, \big[\Xi\cdot{\cal{A}}\big]^m$ etc
 remain the same  in either of  ($\bar e^M\!,\, \bar E_N$), 
($e^M\!,\, E_N$). This is due to the `diagonality' of
$e^\mu\!=\!\bar e^\mu$  $E_n\!=\!\bar E_n$ and survives the 
generalized excitation of $\bar e^\mu$ to
$e^\mu\!\!=\!\!e^\mu_{\bar\mu}dx^{\bar\mu}$, 
etc., to be introduced later-on by  (\ref{gtilt}). 
This tilt invariance can simplify 
calculations  considerably; its members  also include volume elements like
$\bar\varepsilon\!=\!\varepsilon$  and derivations
from $\bar E_n=E_n$,
but  {\em not}  ordinary partial derivatives from $\bar E_\mu\neq E_\mu$.
For the latter kind, by excitation of $\partial_\mu$
under the tilt of the frames in (\ref{tilt}), 
a  rigorous gravito-EW `minimal-coupling'  rule can be uncovered as 
\be
\bar E_\mu
\rightarrow E_\mu=\bar E_\mu-g\big[\Xi\cdot{\cal{A}}\big]_\mu
\Longrightarrow \;\;\;\partial_\mu
\rightarrow\;\;\partial_\mu-g(\xi_i{\cal{A}}^i_{\mu}\sin{\theta}+
\xi_4{\cal{A}}^4_{\mu}\cos{\theta})
  \ .
\label{ttcom}
\ee

To proceed with the calculation of the  ${\cal L}_{\;{\rm HEC}}$ Lagrangian,
we first note that we have not yet arrived at
${\cal M}^8(e^A, \gamma^A_{\;\;B})$ of (\ref{m8}), because we have not
yet excited  the connection. Accordingly, we will employ 
an asterisk $*$ on our not-yet-excited intermediate
${\gamma^*}^M_{\;\;N}$, which, of course, has changed anyway
from the $\bar\gamma^M_{\;\;N}$ value,
due to the $\delta\bar e^A$ exitation, so its Christofel  
part  is identical to the
one involved in the standard Kaluza ansatz.
The calculation towards the intermediate 
${\cal L}^\ast_{\;{\rm HEC}}\!=\!{\cal R}^\ast$  
(modulo surface terms) involves  the basic preliminary result
\be
de^m=g\big[\Xi\cdot{\cal{F}}\big]^m\!\!
-\frac{1}{2{\rm L}_o}\delta^m_i\epsilon^i_{\;jk}e^j\wedge e^k,\;\;\;\;
\big[\Xi\cdot{\cal{F}}\big]^m:=
\Xi^m_i{\cal F}^i\sin{\theta}+\Xi^m_4\cos{\theta}
{\cal F}^4,
\label{eF}
\ee
with the gauge-field strength ${\cal{F}}$ as defined below.
Its  kinetic term emerges  in ${\cal L}^\ast_{\;{\rm HEC}}$ 
as in the standard treatment, while the rest, formally included in
$[{\rm GR+GEW\;terms}]$ sector, relates to gravity and
torsion. In view of the $\bar {\cal R}\!=\!0$  result  in $\bar{\cal M}^8$
there will be no  effective cosmological constant 
from reduction to 4 spacetime dimensions. These aspects of the
\be
{\cal L}^\ast_{\;{\rm HEC}}=
[{\rm GR+GEW\;terms}]-\frac{1}{2}\kappa^2{\cal F}^2,
\;\;\;\;\;\;\;\;
{\cal F}^I:=d{\cal A}^I+\frac{1}{2}g\sin{\theta} \;\d_i^I\e^i_{jk} {\cal A}^j
\wedge{\cal A}^k,
\label{lagr}
\ee
Lagrangian will  not be influenced by the mentioned  generalization in (\ref{gtilt}),
and our remarks following (\ref{killinglengths}) do apply, of course,  to
the gauge-invariant environment  established by (\ref{lagr}).
The  slicing angle $\theta$ therein  is fully redundant
(with a trivial re-definition of  $g\sin{\theta}$) 
and we could have dispensed with it  
{\em and} the Killing vectors altogether. In fact,
it would have been easier to arrive at  (\ref{lagr})
by simply employing,  instead of  ($e^M$, $E_N$), a left-invariant
frame. For that, we could have simply used 
$L_o\delta_I^m{\cal A}^I$
instead of  $\big[\Xi\cdot{\cal A}\big]^m$  in (\ref{tilt}), and then
proceed as usual to verify the claim.
We conclude that, as long as the gauge symmetry is respected,
any particular slicing of the torus is as good as any other, so the slicing
angle ${\theta}$ must drop out in a symmetric environment,
and it does. Indeed, the
$\theta$-dependence of the
$ {\rm L}_o/\!\sin{\theta}$ and ${\rm L}_o/\!\cos{\theta}$   radii in the 
orthonormality relations between the Killing vectors in (\ref{killinglengths}) 
works in conjunction with the standard  choice   in (\ref{dot}) for the
dependence of $\big[\Xi\cdot{\cal{A}}\big]$ on $\theta$,
so the slicing angle  is precisely canceled out.     
However, this  seemingly `useless'  involvement of $\theta$ 
will prove crucial and irreplaceable for the implementation 
of the upcoming gauge-symmetry breaking by the KA2 approach, 
as we'll see in the next section.


\no
\section{Excitation of the torsion to arrive at
${\cal M}^8(e^A, \gamma^A_{\;\;B})$}

For completion of the KA2 approach in (\ref{m8}), our last step involves
the excitation of the torsion
to $ T^M\!=\!\bar T^M\!+\!\delta\bar T^M$ and the induced 
$ K^M_{\;\;N}\!=\!\bar K^M_{\;\;N}\!+\!\delta\bar K^M_{\;\;N}$.
Both excitations are linear in ${\cal A}^I$ 
(as with (\ref{tilt}) for $\delta\bar e^M$)
and  linearly related among themselves (by (\ref{torsion}) etc) as
\be
\delta\bar K_{MNP}=-\frac{1}{2}
\big{(}\delta\bar T_{MNP}+\delta\bar T_{NPM}-\delta\bar T_{PMN}\big{)} 
\ .
\label{kt}
\ee
However, before we proceed with the calculation of this
(proportional to ${\cal A}^I$
and scaled by the mentioned ${{\rm L}_1}$)
$\delta\bar T^M$, we must make sure that this
excitation of  the connection  is  indeed independent, 
namely that the field-content
of ${\cal A}^I$  has not  been already exhausted towards the 
$\delta\bar e^M$ tilt of the frames
in (\ref{tilt}). As we'll see in the last section, 
unspent  degrees of freedom  in ${\cal A}^I$ do survive in this 
case and they are {\em precisely} enough 
to accommodate the transverse degrees of freedom of the EW gauge bosons.
We can easily find that {\em any} 
general  connection ${\gamma}^M_{\;\;N}$, enlarged 
to  ${\gamma}^M_{\;\;N}\!+\!K^M_{\;\;N}$, induces the
$K^{MPN}K_{NPM}- K^{MP}_{\;\;\;\;\;M} K^N_{\;\;PN}$
contribution (modulo surface terms)
to an accordingly enlarged  ${\cal L}_{\;{\rm HEC}}$ Lagrangian.
Thus, when the intermediate 
connection ${\gamma^*}^M_{\;\;N}$ 
is excited  to
$\gamma^M_{\;\;N}\!=\!{\gamma^*}^M_{\;\;N}+\delta\bar K^M_{\;\;N}$, 
the intermediate ${\cal L}^\ast_{\;{\rm HEC}}$
in (\ref{lagr}) will  be accordingly elevated to final form
with quadratic-in-$\delta\bar K$ terms, as
\be
{\cal L}_{\;{\rm HEC}}=[{\rm GR+GEW\;terms}]-\frac{1}{2}\kappa^2{\cal F}^2+
\delta \bar K^{APB}\,\delta\bar K_{BPA}-
\delta\bar K^{MP}_{\;\;\;\;\;M}\,\delta\bar K^N_{\;\;PN}.
\label{flagr}
\ee
Due to the implicit presence of quadratic-in-${\cal A}^I$
terms (in the two last ones on the r.h.s.), we have already lost
the $SU(2)\times\!U(1)$ left invariance which had previously covered 
the entire  ${\cal L}^\ast_{\;{\rm HE}}$
in (\ref{lagr}), so gauge-symmetry breaking has already occurred  in 
the ${\cal L}_{\;{\rm HEC}}$  of (\ref{flagr}), as a result of the
excitation $\delta\bar T^M$ of the connection.

To replace $\sim$ with precise  equality in 
$\delta\bar T^M\!=\!\frac{1}{2}\delta\bar T^M_{\;\;NP}e^N\wedge e^P\sim
{{\rm L}_1^{-1}}{\cal A}^I$, we note that the missing tensorial factor on the r.h.s.
must: depend only 
on the Killing vectors $\Xi_J$, have exactly one free
group-index $I$ (to saturate  the free-one on ${\cal A}^I$) and
balance the rest of the free indices in   
that relation. To accomplish that, we must exploit  
the  already-installed breaking of 
gauge invariance in (\ref{flagr}), in the sense that  there exists
an unknown but  {\em specific} angle $\theta_W$, by which 
the $S^3\times S^1$ torus can be viewed as already sliced.
By  fixing (in-retrospect agreement with standard convention) the 
$I\!=\!3$ direction, we can introduce a fixed mixing as
$\Xi_{(W)}^b\sim(\Xi_3^b\sin{\theta_W}+
\Xi_4^b\cos{\theta_W})$ without loss of generality. This $\Xi_{(W)}^b$
times a $\Xi_I^p$ for the required free index $I$ (and antisymmetry
for torsion)
accommodates fully and  precisely all the
requirements on our  tensorial factor as $\Xi_{(W)}^{[b}\Xi_I^{p]}$, 
so the final result is {\em unique} as
\be
{\delta T}^\rho= \frac{g}{\rm L_1}\,\eta^{\rho\sigma}\eta_{bm}\eta_{pn}
\Xi_{(W)}^{[b}\Xi_I^{p]}{\cal A}^I_\sigma\, e^m\wedge e^n,\;\;\;\;
\Xi_{(W)}^b:=\frac{1}{\sqrt{2}{\,\rm L_o}}\left(\Xi_3^b\sin{\theta_W}+
\Xi_4^b\cos{\theta_W}\right).
\label{t}
\ee
Thus, by (\ref{kt}), the only non-vanishing independent components of 
$\delta\bar T^M$ and $\delta\bar K^M_{\;\;N}$  are
\be
 \frac{1}{2}\delta\bar T^{\mu bp}=-\delta\bar K^{\mu bp}=\delta\bar K^{bp\mu}=
\frac{g}{{\rm L_o}{\rm L}_1}\eta^{\mu\nu}\Xi_{(W)}^{[b}\Xi_I^{p]}{\cal A}^I_\nu \ .
\label{ktc}
\ee

\no
To find explicitly the  mass term 
already present in  (\ref{flagr}), we may re-write the latter as
\be
 {\cal L}_{\;{\rm HEC}}=[{\rm GR+GEW\;terms}] - 
\frac{\kappa^2}{2}{\cal F}^2-
\kappa^2M_{IJ}{\cal{A}}^I_\mu{\cal{A}}^J_\nu\eta^{\mu\nu}
 \ ,
\label{lagm}
\ee
wherefrom, by the  tracelessness of
contorsion  from (\ref{ktc}), we can read-out the identification
\be
\kappa^2M_{IJ}{\cal{A}}^I_\mu{\cal{A}}^J_\nu\eta^{\mu\nu}=
 - \delta K^{APB}\delta K_{BPA}
 \ .
\label{mk}
\ee
The straightforward  substitution of (\ref{ktc})  in (\ref{mk})  
quantifies the mass matrix etc., as
\be
M_{IJ}=({\rm L}_o{\rm L}_1)^{-2} \left[(\Xi_{(W)})^2\,\Xi_I^b\Xi_J^p\,
\eta_{bp}-(\Xi_{(W)}\cdot\Xi)_I(\Xi_{(W)}\cdot\Xi)_J \right], 
\label{mm}
\ee
\be
 (\Xi_{(W)}\cdot\Xi)_I:=\Xi_{(W)}^b\Xi_I^p\,\eta_{bp} =
\frac{{\rm L}_o}{\sqrt{2}}\left(\frac{\sin{\theta_W}}{\sin^2{\theta}}\delta_{3I} + 
\frac{\cos{\theta_W}}{\cos^2{\theta}}\delta_{4I}\right),
\label{xx}
\ee
where, having used the
orthonormality relations from (\ref{killinglengths}),
we must now set $\theta=\theta_W$. With  
$(\Xi_{(W)})^2\!=\!1$, as normalized in (\ref{t}),  
we may express 
the mass-term in (\ref{mm}) as  
\be
M_{IJ}= 
({\rm L}_1\sin{\theta_W})^{-2}\left[
\eta_{ij}\delta^i_I\delta^j_J-\frac{1}{2}
\left(\delta^3_I\delta^3_J+\tan^2{\theta_W}\delta^4_I\delta^4_J 
+\tan{\theta_W}(\delta^3_I\delta^4_J +\delta^4_I\delta^3_J)\right)\right],
\label{mm1}
\ee
or, equivalently, in the more conventional  matrix notation, as
\be
M_{IJ}=({\rm L}_1\sin{\theta_W})^{-2}
\left(\begin{array}{ccccc} 1&{\;} & 0&0&0 \\ 
0&{\;}&1&0&0\\0&{\;}&0&\frac{1}{2}&-\frac{1}{2}
\tan{\theta_W}\\0&{\;}&0&-\frac{1}{2}\tan{\theta_W}&
\frac{1}{2}\tan^2{\theta_W}
\end{array} \right)          
 \ .
\label{mm2}
\ee
Either  
\be
\Delta^I_{\;\;J}=
\left(  \begin{array}{ccccc} 1&{\;} & 0&0&0 \\
 0&{\;}&1&0&0\\0&{\;}&0&-\cos{\theta_W}& 
\sin{\theta_W}\\0&{\;}&0&+\sin{\theta_W}&\cos{\theta_W}
\end{array} \right)  \;{\rm {or}}\;\;
\left(  \begin{array}{cccc} 1/\sqrt{2} & 1/\sqrt{2}&0&0 \\ 
-i/\sqrt{2}&i/\sqrt{2}&0&0\\0&0&-\cos{\theta_W}& \sin{\theta_W}\\
0&0&+\sin{\theta_W}&\cos{\theta_W} \end{array} \right)          
 \ 
\label{diag}
\ee
diagonalizes $M_{IJ}$  to its eigenvalues as
\be
M^I_{\;\;J}=({\rm L}_1\sin{\theta_W})^{-2}
\left(  \begin{array}{ccccc} 1&{\;} & 0&0&0 \\ 0&{\;}
&1&0&0\\0&{\;}&0&\frac{1}{2}(\cos\theta_W)^{-2}& 0\\0&{\;}&0&0&0
\end{array} \right)=        
\left(  \begin{array}{cccc} m^2_W & 0&0&0 
\\ 0&m^2_W&0&0\\0&0&\frac{1}{2}m^2_Z& 0\\0&0&0&0
\end{array} \right)   
 \ , 
\label{diagm}
\ee
so the gauge-boson  masses are post-dicted as
 $m_W\!=\!({\rm L}_1\sin{\theta_W})^{-1}$, 
$m_Z\!=\!({\rm L}_1\sin{\theta_W}\cos{\theta_W})^{-1}$,  and 
the $\rho\!:=\!m^2_W/( m_Z\cos{\theta_W})^2$ parameter as $\rho\!=\!1$.
For the physical gauge bosons, actually read off (\ref{diag}) as
$W^\pm\!=\!({\cal{A}}^1 \mp i{\cal{A}}^2)/\sqrt{2}$,
$Z\!=\!-\cos{\theta_W}{\cal{A}}^3\!+\!\sin{\theta_W}{\cal{A}}^4$,
$B\!=\!\sin{\theta_W}{\cal{A}}^3\!+\!\cos{\theta_W}{\cal{A}}^4$,
the  ${\cal L}_{\;{\rm HEC}}$
of (\ref{lagm}) takes on its standard expression, 
with the mass term therein as
\be
\kappa^2M_{IJ}{\cal{A}}^I_\a{\cal{A}}^J_\b\d^{\a\b}=
\kappa^2\big(m^2_WW^+W^-+\frac{1}{2}m^2_ZZ^2\big)
 \ .
\label{massterm}
\ee

\section{Discussion}

In spite of its simplicity, the  KA2
approach by (\ref{m8}) vs the standard  (\ref{m8s})
offers surprisingly far-reaching results and relevance to 
fundamental issues, hitherto
unrelated. We will briefly expand
toward some of them,  after some clarifications and pending completions.
As anticipated (in view of the  $\bar {\cal R}\!=\!0$ 
result  in $\bar{\cal M}^8$), in order to have a  general
curvature scalar ${\cal R}$ present in the
$[{\rm GR+GEW\;terms}]$ general relativistic and gravito-EW sector in (\ref{lagm}),
we must allow for a generalized excitation
of  $\bar e^\mu$ to  $e^\mu$ in (\ref{tilt}),
now re-defined in terms of the vierbeins  
$e^\mu_{\bar\rho}$ and inverse $ E_\nu^{\bar\rho}$ 
(namely with $e^\mu_{\bar\rho} E_\nu^{\bar\rho}=
\delta^\mu_\nu$) as
\be
 e^A=e^\mu_{\bar\mu}\,dx^{\bar\mu}\delta_\mu^A+
\left(\bar e^m+
g\big[\Xi\cdot{\cal{A}}\big]^m\right)\delta_m^A
 \;\;\; \Longleftrightarrow\;\;\;
 {E_B}=\left( E_\nu^{\bar\nu}\,\partial_{\bar\nu}-
g\big[\Xi\cdot{\cal{A}}\big]_\nu\right)\delta^\nu_B+
\bar E_n\delta^n_B
  \ .
\label{gtilt}
\ee
When we set out for the simplest non-trivial implementation 
of the KA2 approach, we initially anticipated to utilize one of the
$\bar{\cal M}^7\!=\!\bar{\cal B}^4_{\rm T}\times \IR^3$ or
$\bar{\cal M}^7\!=\!\bar{\cal B}^4_{\rm M}\times \IR^3$
ground states, instead of the finally employed 
$\bar{\cal M}^8\!=\!\bar{\cal B}^4_{\rm M}\times S^1\times \IR^3$ 
in (\ref{m8}). The $S^1$ factor therein is not
merely a `spectator', but it is rather imposed by
{\em augmendability} under  KA2,  as we'll see.
By general considerations \cite{duff}, the 7-dimensional 
${\cal M}^7$ proper vacuum would have involved a total of $7(7\!-\!3)/2\!=\!14$
independent states in its ${\cal L}_{\;{\rm HEC}}$; 
the  piecemeal count of the 2 graviton states in ${\cal M}^7$
plus the $SU(2)$  scalar and {\em massless} gauge boson states as 
$2\!+\!3(3\!+\!1)/2\!+\!2\!\cdot\!3$  would have again 
given us precisely 14.
This means that
there would be no field-content left in the corresponding $\cal A^I$
for an independent variation of the torsion. Thus,
for our ${\cal M}^7$ cases, we would  end-up with
a gauge-invariant ${\cal L}_{\;{\rm HEC}}$ and no mass-term, 
hence with a KA2 approach redundant to the standard one. 
A  non-redundant KA2 is  achieved
with the minimally augmented 
${\cal C}^5\!=\!{\cal B}^4_{\rm M}\times S^1$ for the 
$\bar{\cal M}^8\!=\!\bar{\cal B}^4_{\rm M}\times S^1\times \IR^3$
ground state in (\ref{m8}).
Indeed, in this case, the general count of 
$8(8\!-\!3)/2\!=\!20$ states for ${\cal M}^8$
is not quite matched by the piecemeal count of  states, 
{\em again} for massless (now $SU(2)\times U(1)$) gauge bosons,
because  $2\!+\!(3(3\!+\!1)/2\!+\!1)\!+\!2\!\cdot\!(3\!+\!1)\!=\!17$.
The surviving 3 degrees of freedom in $\cal A^I$ 
have been precisely enough to generate 
(by independent excitation of the torsion) the
3 transverse states of the mass term (\ref{massterm}) in  
the  Lagrangian (\ref{lagm}). 
Thus, the notion of {\em augmendability} by  KA2 emerges as  
complementary to the requirement for spontaneous compactification. 
As a result, ${\cal C}^5\!=\!{\cal B}^4_{\rm M}\times S^1$  is 
augmendable  under KA2, but the ${\cal C}^4\!=\!{\cal B}^4_{\rm IX}$  is not.

We are now better equipped to sum-up our results as follows. \newline
(i) The  gravitational and electroweak sectors have 
emerged in elegant {\em hierarchy} in a
${\cal L}_{\;{\rm HEC}}$ of the form (\ref{lagm}), in terms of the
(identifiable as gravitational) coupling $\kappa$ and the EW scale ${\rm L_1}$. 
The gravitational interaction is an effective one at scales
${\rm L}\!>\!>\!{\rm L_o}$ and its frames $e^\mu$, 
as defined by (\ref{gtilt}), are {\em in principle} calculable via  
Einstein's equations. The latter will follow from any  Lagrangian 
of the type (\ref{lagm}), after the fashion of  `electrovac equations' 
and a `minimal-coupling' rule, now elevated to  
gravito-EW vacuum equations and (\ref{ttcom}), respectively.
\newline
(ii) The {\em higgsless}  emergence of the EW gauge-boson masses
is fully calculable by
(\ref{mm}-\ref{massterm}), although the numerical value of
$\theta_W$ would be calculable only with  the employment of the
$\bar{\cal B}^4_{\rm T}$ of the Taub string
(instead of the $\bar{\cal B}^4_{\rm M}$) in (\ref{m8}), via the 
set of the radii of its squashed $S^3$, the
$(1/\sqrt{3},1/\sqrt{3},\sqrt{\pi/2-1})$ 
in units of ${\rm L_o}$  \cite{bat1}.
The mass term  in (\ref{lagm}) has been
produced  by the geometry
via excitation of the effective torsion, actually the
only geometric element which could
protect against  mathematical singularities, if masses were to be
added by-hand.
\newline
(iii) By KA2, if mathematical singularities (but not {\em physical} ones) 
were to be excluded from physical spacetime, 
the  fundamental  r\^ole of gravity in the dynamics of vacuum geometry is being 
conceded to the remaining fundamental interactions. Gravity  does  retain 
all its geometric aspects, but
the dimensionful coupling $\kappa$ 
is now its only relation to Planck scale.
At or close to that scale, where everything is 
{\em actually} part of a true proper vacuum,
the meaning of a gravitational coupling is empty anyway.
At the intermediate  regime, where all other interactions are quantized 
(say, very widely around ${\rm L_1}$), gravity would again be 
in a ${\rm L}\!>\!>\!{\rm L_o}$
environment, so it would  remain classical there,
as in ordinary 4D classical  regime.
It would then follow that gravity can only stand as an 
{\em effective} interaction or classical field in 4D, 
and as such it would have to be excluded from quantization.

Thus, by our findings via the KA2 approach, we may conjecture
an augmentable ${\cal C}^n$ to adjoin
the strong interaction towards  a standard model 2,
which already includes gravity. 
In view of the reasonable  $n\leq 7$ requirement, this would also
offer the option of an analogous re-orientation in supergravity
\cite{duff}.

\vskip .1in
\no
I am grateful to the ever-inspiring contributors to our {\em milieux}
for ideas, prompts and ways to go after the awesome taste of pure marble-stone soup!


\end{document}